\documentclass[aps,twocolumn,superscriptaddress]{revtex4-1}     % ,reprint,groupedaddress
\usepackage{graphicx}
\usepackage{amsmath,amssymb}
\usepackage{dsfont}
\usepackage{bm}
\usepackage{MnSymbol, marvosym, wasysym}

\renewcommand{\vec}[1]{\boldsymbol{#1}}
\newcommand{\w}{{\omega}}

\newcommand{\pd}{{\phantom{\dagger}}}
\newcommand{\bs}[1]{\boldsymbol{#1}}

\newcommand{\ie}{{\it i.e.},\ }
\def\eg{\emph{e.g.}\ }
\def\ea{\emph{et al.}}
\def\cf{\emph{c.f.}\ }

\usepackage{color}
\definecolor{darkgreen}{rgb}{0.10, 0.65, 0.10}

\newcommand{\sr}[1]{{\color{black}#1}}

\begin{document}

%\title{Fate of local detection of topological insulators}
\title{Impurity bound states as detectors of topological band structures revisited}
%\title{Impurity bound states as detectors of topological band structures}

\author{Seydou-Samba Diop}
\affiliation{School of Physics, University of Melbourne, Parkville, VIC 3010, Australia}
\affiliation{D\'{e}partement des Sciences de la Mati\`{e}re, ENS de Lyon, 69007 Lyon, France}
\author{Lars Fritz}
\affiliation{Institute for Theoretical Physics, Utrecht University, Princetonplein 5, 3584 CC Utrecht, Netherlands}
\author{Matthias Vojta}
\affiliation{Institut f\"{u}r Theoretische Physik, Technische Universit\"{a}t Dresden, 01062 Dresden, Germany}
\author{Stephan Rachel}
\affiliation{School of Physics, University of Melbourne, Parkville, VIC 3010, Australia}

\date{\today}

%%%%%%%%%%%%%%%%%%%%%%%%%%%%%%%%%%%%%%%%%%%%%%%%%%%%%%%%%%%%%%%%%%%%%%%%%%%

\begin{abstract}
Band structures of topological insulators are characterized by \textit{non-local} topological invariants. Consequently, proposals for the experimental detection using \textit{local} probes are rare.
A recent paper [Slager {\it et al.}, Phys.\ Rev.\ B {\bf 92}, 085126 (2015)] has argued, based on theoretical results for a particular class of models, that insulators with topologically trivial and non-trivial band structures in two space dimensions display a qualitatively different response to point-like impurities.
Here we present a comprehensive investigation of the impurity response of a large set of models of non-interacting electrons on the honeycomb lattice, driven insulating by either broken inversion, broken time reversal, broken $C_3$, or broken translation symmetry.
%, or by application of magnetic field or inhomogeneous strain. 
These cases include Hofstadter bands, strain-induced pseudo-Landau levels and higher-order topological insulators.
Our results confirm that for hopping models respecting the lattice symmetries, the response to a single impurity can indeed distinguish between trivial and non-trivial band topology. However, for modulated or inhomogeneous host systems we find that trivial states of matter can display an impurity response akin to that of topologically {\it non-trivial} states, and thus the diagnostic fails.
\end{abstract}

\maketitle

%%%%%%%%%%%%%%%%%%%%%%%%%%%%%%%%%%%%%%%%%%%%%%%%%%%%%%%%%%%%%%%%%%%%%%%%%%%
%%%%%%%%%%%%%%%%%%%%%%%%%%%%%%%%%%%%%%%%%%%%%%%%%%%%%%%%%%%%%%%%%%%%%%%%%%%
%%%%%%%%%%%%%%%%%%%%%%%%%%%%%%%%%%%%%%%%%%%%%%%%%%%%%%%%%%%%%%%%%%%%%%%%%%%

\section{Introduction}
\label{sec:intro}

Topological insulators (TIs) constitute one of the most active fields of contemporary condensed matter research\,\cite{qi-11rmp1057,hasan-10rmp3045,bernevig13}. Their theoretical prediction\,\cite{kane-05prl146802,kane-05prl226801,bernevig-06prl106802,bernevig-06s1757,moore-07prb121306,roy09prb195322,fu-07prl106803} and subsequent experimental realization\,\cite{koenig-07s766,hsieh-08n970} -- both in two and three spatial dimensions -- also led to the discovery of a variety of other topological phases\,\cite{wan-11prb205101,burkov-11prl127205,fu11prl106802,bradlyn-17n257,rachel18rpp116501,benalcazar-17s61} and motivated the concept of symmetry-protected topological phases\,\cite{pollmann-10prb064439,wen12prb085103,chen-12s1604} as well as the topological classification of all free-fermion states\,\cite{schnyder-08prb195125,kitaev09,slager-13np98,kruthoff-17prx041069}. The continued interest in these novel states of matter is due not only to their fundamental importance, but also to their application prospects, e.g., for low-power electronics\,\cite{collins-18n390} thanks to dissipationless edge and surface transport.

TIs are characterized by an insulating bulk and metallic edge or surface states. Equivalently, one can define them by means of topological invariants which are calculated from their quantum-mechanical bulk wavefunctions. This equivalence is referred to as {\it bulk--boundary correspondence}. The metallic boundary states are protected against disorder and other small perturbations as long as the protecting symmetry is preserved.
%In three dimensions (3D), the topological field theory of TIs\,\cite{qi-08prb195424} predicts axion electrodynamics\,\cite{wu-16s1124} and the topological magnetoelectric effect\,\cite{dziom-17nc15197}.

Experimentally, TIs are typically identified via their boundary states, either by transport measurements, \eg using Hall bar geometries\,\cite{koenig-07s766}, or by spectroscopic imaging using angle-resolved photoemission spectroscopy (ARPES)\,\cite{hsieh-08n970} or scanning tunneling microscopy and spectroscopy (STM/STS)\,\cite{reis-17s287,collins-18n390,pauly-16acsn3995}. Since boundary states can also have a non-topological origin, such experiments need to be combined with a theoretical analysis in order to give conclusive evidence. Given that topology is a global property, an unambiguous detection using \emph{local} observables is not possible in principle; however, it is of practical interest to develop local indicators for topological states of matter. In this context, the behavior near defects, \ie impurities, edges and dislocations has been investigated for different topological systems \,\cite{teo-10prb115120,lu-11njp103016,kimme-16prb035134,black-schaffer-12prb115433,ran-09np298,slager-14prb241403}.

A recent paper\,\cite{slager-15prb} argued that the spectral response to a single impurity placed in an otherwise clean system of weakly interacting electrons can serve as a clear-cut signature of non-trivial topology:
Introducing a potential scattering impurity into the bulk of an insulator may lead to electronic states bound to the impurity whose energy is located outside the bulk bands. For the Bernevig-Hughes-Zhang (BHZ) model\,\cite{bernevig-06s1757}, Ref.\,\onlinecite{slager-15prb} deduced that impurities of arbitrary strength always induce a bound state energetically located in the bulk gap for host states with non-trivial topology, but not so for topologically trivial states. In the latter case, in-gap bound states are absent for strong impurities.
% sentence simplified
This was shown to apply to impurities with both codimension $1$ (\ie defect planes in 3D and lines in 2D) and codimension $2$ (\ie defect lines in 3D and points in 2D) and could be related to the presence of zeroes of the local host Green's function in the gap. 
Intuitively, a strong impurity expels electrons and acts as a topologically trivial region, thus inducing a topological ``edge'' (i.e. impurity) state if the surrounding bulk is topologically non-trivial. Such impurity states being absent for trivial bulk states is the key finding of Ref.\,\onlinecite{slager-15prb}.
We note that the different impurity response for topological and trivial phases was discussed earlier for the Kane-Mele model\,\cite{gonzalez-12prb115327}; the codimension-1 case was also investigated in Ref.\,\cite{pinon-19arXiv1906.08268}.

In the present paper, we address the key question how general the concept of impurity bound states as detectors of topological band structures actually is. Given the tremendous progress in the artificial engineering of 2D lattices, we primarily focus on the experimentally relevant case of point-like impurities in 2D systems. We consider lattice models of non-interacting electrons, realizing a variety of insulating phases, both topological and non-topological, and probe their spectral response to either a site or a bond impurity, see Fig.~\ref{fig:models}. For simplicity, we restrict our attention to spinless electrons or, equivalently, situations without spin mixing. We note that our results will thus also apply to spinful extensions such as Quantum spin Hall insulators\,\cite{gonzalez-12prb115327}.

\subsection{Summary of results}
\label{sec:results}

For Hamiltonians preserving the lattice symmetries we confirm the scenario put forward in Ref.\,\onlinecite{slager-15prb}:
For topological states characterized by a finite Chern number, \ie Chern or quantum Hall insulators, we find one (or several) in-gap bound state(s) for arbitrary impurity strength, regardless of the type of impurity [see \eg Fig.\,\ref{fig:haldane_impurity_spectra}\,(a) below]. Conversely, for topologically trivial states there are no in-gap bound states for large impurity strength [see \eg Fig.\,\ref{fig:haldane_impurity_spectra}\,(c)].

In contrast, for Hamiltonians breaking lattice symmetries via anisotropic or modulated hopping matrix elements, the situation is different. We find that many of such models feature an impurity response akin to that of a topological phase, i.e., impurity bound states occur for arbitrary impurity strength. However, the origin of those bound states is clearly non-topological, as it can be traced back to the behavior of isolated \sr{oligomers (\ie dimers, trimers, tetramers, pentamers, hexamers)}.
Examples for such models include the honeycomb lattice with hopping anisotropies, with Kekul\'e modulation, and with triaxial strain pattern. While some of them can be related to higher-order topological insulators \cite{benalcazar-17s61}, their impurity response is unrelated to this fact.
%
%Our results thus reveal that there is a large family of topologically trivial systems which feature an impurity response suggesting topologically non-trivial behavior, i.e., the diagnostic proposed in Ref.\,\onlinecite{slager-15prb} delivers false positive results.
%
%Our results thus reveal that there is a large family \sr{of systems (including topologically trivial ones)} which
%
Our results thus reveal that there is a large family of systems which feature an impurity response suggesting topologically non-trivial behavior, regardless of whether they are in a topological phase or not. That is, the diagnostic proposed in Ref.\,\onlinecite{slager-15prb} delivers false positive results.

\subsection{Outline}

The remainder of the paper is organized as follows: %firstly we review the main idea of Ref.\,\cite{slager-15prb} and the implications thereof in Sec.\,\ref{sec:slager}. Then
Sec.\,\ref{sec:haldane} introduces the formalism used in the paper and discusses the Haldane and Semenoff insulators, representing topologically non-trivial and trivial states of matter, respectively, as the spinless analogues on the honeycomb lattice of the two distinct phases of the BHZ model discussed in Ref.\,\onlinecite{slager-15prb}.
Sec.\,\ref{sec:tatbtc-model} deals with the gapped phases of the anisotropic honeycomb lattice breaking $C_3$ rotation symmetry. We will relate their behavior to that of the one-dimensional Su--Schrieffer--Heeger (SSH) model and its spectral response to impurities.
Sec.\,\ref{sec:PLL} is devoted to the honeycomb lattice under triaxial strain leading to pseudo-Landau levels in the spectrum.
%In contrast to the discussion in Ref.\,\cite{rachel-16prl-116}, we consider the {\it exact} version\,\cite{rachel-16prl-117} of strained Landau levels having the advantage that finite-size and edge effects are absent.
The subsequent Sec.\,\ref{sec:hofstadter} discusses the impurity response of Landau levels due to an orbital magnetic field (aka Hofstadter model) which we compare to that of the strain-induced Landau levels and of the Chern-insulator phase of the Haldane model.
Eventually, Sec.\,\ref{sec:HOTI} elaborates on the impurity response of higher-order topological insulator (HOTI) phases realized by a Kekule or anti-Kekule distortion of the honeycomb lattice.
The central results of the different models are discussed in a broader context in Sec.\,\ref{sec:discussion}. We end with a summary in Sec.\,\ref{sec:summ}.

%%%%%%%%%%%%%%%%%%%%%%%%%%%%%%%%%%%%%%%%%%%%%%%%%%%%%%%%%%%%%%%%%%%%%%%%%%%
%%%%%%%%%%%%%%%%%%%%%%%%%%%%%%%%%%%%%%%%%%%%%%%%%%%%%%%%%%%%%%%%%%%%%%%%%%%
%%%%%%%%%%%%%%%%%%%%%%%%%%%%%%%%%%%%%%%%%%%%%%%%%%%%%%%%%%%%%%%%%%%%%%%%%%%

\section{Haldane and Semenoff insulators}
\label{sec:haldane}

%%%%%%%%%%%%%%%%%%%%%%%%%%%%%%%%%%%%%%%%%%%%%%
\begin{figure}[t!]
\begin{center}
\includegraphics{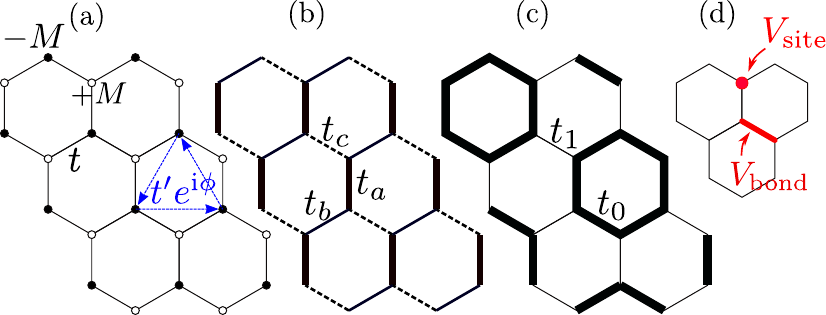}
\caption{
Some of the tight-binding models on the honeycomb lattice discussed in this paper.
(a) Haldane model \eqref{eq:haldane_hamiltonian} involving real nearest-neighbor hopping with amplitude $t$, the staggered Semenoff sublattice potential with onsite energies $-M$ ($+M$) on sublattice $A$ ($B$) and the complex-valued second-neighbor hopping $t' e^{i\phi}$ shown for one triangle of $A$ sites.
(b) Anisotropic honeycomb model \eqref{eq:tatbtc} with different hopping amplitudes $t_a$, $t_b$ and $t_c$ for the three different nearest-neighbor hopping directions. 
(c) (Anti-)Kekul\'e hopping modulation \eqref{eq:HOTI} involving $t_0$ and $t_1$ corresponding to the plaquette anisotropy of the higher-order topological insulator phase.
(d) Site and bond impurities, respectively.}
\label{fig:models}
\end{center}
\end{figure}
%%%%%%%%%%%%%%%%%%%%%%%%%%%%%%%%%%%%%%%%%%%%%%

Haldane's Chern insulator model\,\cite{haldane-88prl} is described by the following tight-binding Hamiltonian:
\begin{equation} \label{eq:haldane_hamiltonian}
H_0 = -t \sum_{\langle i, j \rangle} c_{i}^{\dagger} c_{j} -t' \sum_{\langle \!\langle i, j \rangle \!\rangle} e^{\pm i \phi} c_{i}^{\dagger} c_{j} + M \sum_{i} \xi_{i} c_{i}^{\dagger} c_{i}\,.
\end{equation}
The first term is the real-valued hopping term between first-neighbor sites $\langle i,j \rangle$, responsible for Dirac cones in the dispersion relation known from graphene\,\cite{castroneto-09rmp109}. The second term represents complex-valued hopping term between second neighbors $\langle \!\langle i,j \rangle\! \rangle$, breaking time-reversal symmetry (unless $\phi=0, \pi$). Finally, the third term is a staggered on-site potential, originally introduced by Semenoff\,\cite{semenoff-84prl}, which breaks inversion symmetry and causes an imbalance between the two sublattices of the honeycomb lattice. We set $\xi_i = -1$ ($\xi=+1$) if site $i$ belongs to sublattice A (B), see Fig.\,\ref{fig:models}\,a. For what follows we set $t=1$ and $\phi = \pi/2$ unless noted otherwise.

The model in Eq.~\eqref{eq:haldane_hamiltonian} is a semimetal for $M=t'=0$, with band-touching points at momenta $K$ and $K'$. Haldane's Chern insulator is realized for $t'\neq 0, \phi\not= 0,\pi$ and $M=0$ and the Semenoff insulator for $t'= 0, M \neq 0$. These two insulating phases are topologically distinct. For $\phi = \pi/2$ the system remains in the topological phase (characterized by a Chern number $+1$) even for finite $M$ as long as $M/t'<3\sqrt{3}$. For $M/t'>3\sqrt{3}$ the system is in the trivial phase (with Chern number 0). At the transition,  $M/t'=3\sqrt{3}$, the system is semimetallic, but gapless only at momentum $K$, being different from graphene\,\cite{haldane-88prl}.

A local impurity is added in the unit cell $\vec{r}=\vec{0}$:
\begin{equation}
H_{V} =  c^{\dagger}_{\vec{r}= \vec{0}} V_0\hat{V} c_{\vec{r}= \vec{0}}\,.
\end{equation}
Here, the two-component spinor $c^{\dagger}_{\vec{r}= \vec{0}} = (c^{\dagger}_{\vec{r}= \vec{0},A}, c^{\dagger}_{\vec{r}= \vec{0},B})$ lives in the defect unit cell, the scalar parameter $V_0$ measures the impurity strength, and $\hat{V}$ is a normalized $2 \times 2$ Hermitian matrix describing the type of the impurity.
$\hat{V}$ can be expanded into the set of Pauli matrices $\sigma^{i}$ ($i=1,2,3$) and the unit matrix $\mathds{1}$, acting in sublattice space.
We will consider impurities with different sublattice structure:
\begin{itemize}
 \item $\hat{V}=\frac{1}{2}(\mathds{1} \pm \sigma^{3})$, \ie a site impurity acting on either sublattice A or sublattice B;
 \item $\hat{V}=\sigma^{1}$, \ie a (real-valued) bond impurity;
 \item $\hat{V}=\sigma^{2}$, \ie an imaginary-valued bond impurity;
 \item $\hat{V}=\mathds{1}$, \ie two neighboring site impurities of equal strength;
 \item $\hat{V}=\sigma^{3}$, \ie two neighboring site impurities of opposite strength,
\end{itemize}
with the first two being most important. We note that the point impurities discussed by Slager \ea\,\cite{slager-15prb} for the 2D BHZ model\,\cite{bernevig-06s1757} involve both orbitals of the model and hence correspond to our case of two neighboring \sr{site} impurities, $\hat{V}=\mathds{1}$.

%%%%%%%%%%%%%%%%%%%%%%%%%%%%%%%%%%%%%%%%%%%%%%
\begin{figure}
\begin{center}
\includegraphics[width=\columnwidth]{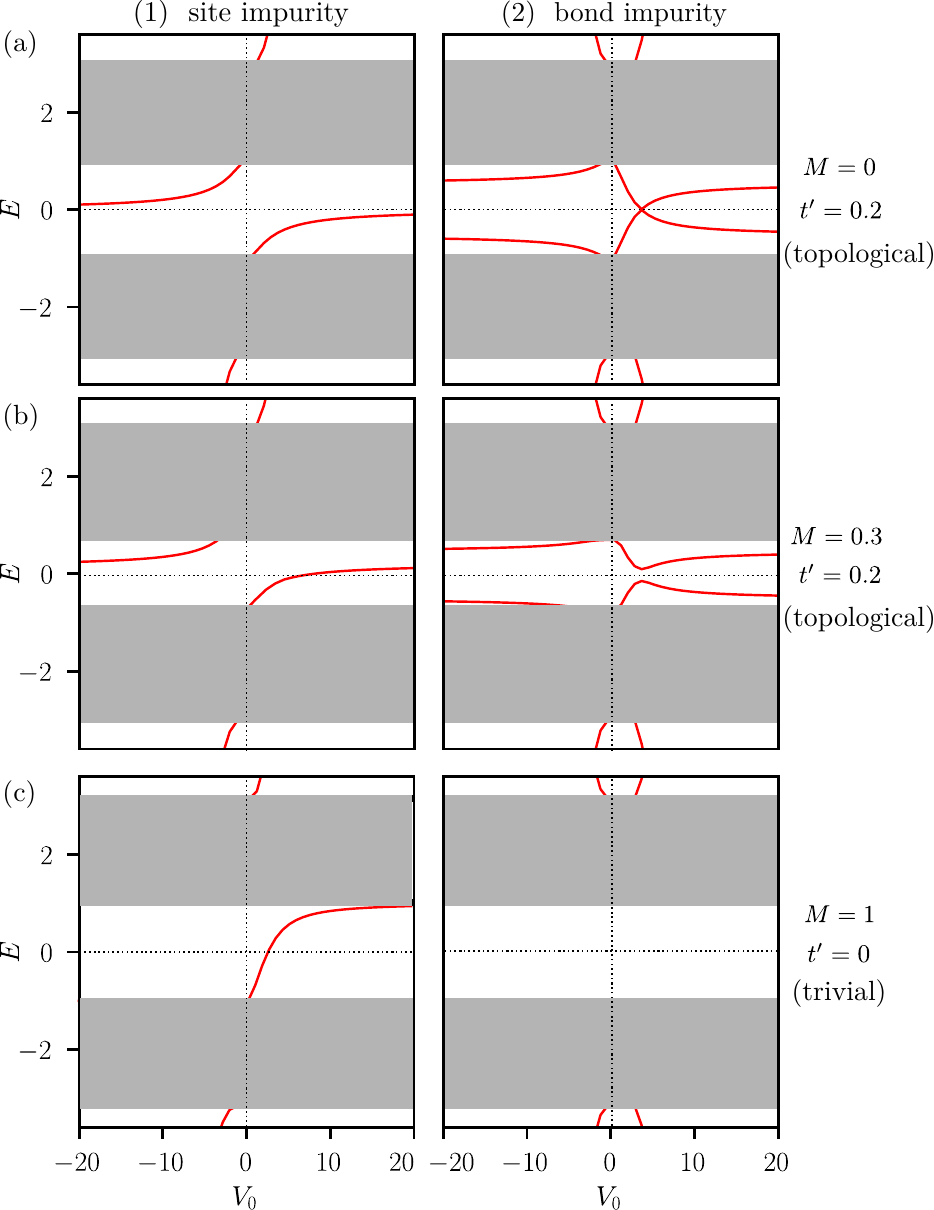}
\caption{
Single-particle levels of the Haldane model with Semenoff term \eqref{eq:haldane_hamiltonian} in the presence of a single impurity of strength $V_0$. Grey regions correspond to the bulk states while impurity bound states are plotted in red. The first column shows results for a site impurity, the second for a bond impurity.
(a) Chern insulator phase with $t'=0.2$ and $M=0$.
(b) Chern insulator phase ($t'=0.2$) in the presence of a finite Semenoff potential $M=0.3$.
(c) Trivial insulating phase ($t'=0$, $M=1$).
In the topological cases (a) and (b) there is one (or two) in-gap bound state(s) present for any $V_0$. In contrast, the trivial insulating case displays an in-gap bound state only in a restricted region of $V_0$ (or none at all).
}
\label{fig:haldane_impurity_spectra}
\end{center}
\end{figure}
%%%%%%%%%%%%%%%%%%%%%%%%%%%%%%%%%%%%%%%%%%%%%%

We diagonalize the total Hamiltonian $H = H_0 + H_{V}$ on a finite lattice of $N_s$ sites with periodic boundary conditions (unless explicitly mentioned otherwise); most figures have been generated with $N_s=400$. The resulting energy levels can be plotted as a function of the impurity strength $V_0$, yielding (almost) continuous and $V_0$-independent bulk bands as well as isolated impurity bound states.
This is shown in Fig.\,\ref{fig:haldane_impurity_spectra} for the model \eqref{eq:haldane_hamiltonian} for three different choices of parameters and two types of impurities.
From the bound-state behavior at large $V_0$ two cases can be clearly distinguished: In the topological Haldane-insulator phase in panels (a,b) in-gap bound states exist for all $V_0$, whereas such bound states are not present in the trivial Semenoff-insulator phase in panel (c). In other words, the existence of in-gap bound states requires fine-tuning in the trivial phase while it is generic in the topological phase; this applies to both a site and a bond impurity. Comparing panels (a) and (b) shows that variations of parameters (here $M$) within the topological phase shifts the bound-state energies within the gap; we note that increasing $M$ also decreases the bulk gap until the topological phase is destroyed.

%
%The transition between the two behaviors can be described as following. Starting from $M=0$ and finite $t'$ (topological phase), the effect of the Semenoff mass is to shift the asymptotic energy of the bound state towards one of the bulk bands. At the same time, it closes the gap between the bands until $M=3\sqrt{3}t'$. By increasing further $M$ the gap reopens into the trivial phase, but the energy of the bound state is already in the band.

%%%%%%%%%%%%%%%%%%%%%%%%%%%%%%%%%%%%%%%%%%%%%%
\begin{figure*}[htbp]
\begin{center}
\includegraphics[width=0.98\textwidth]{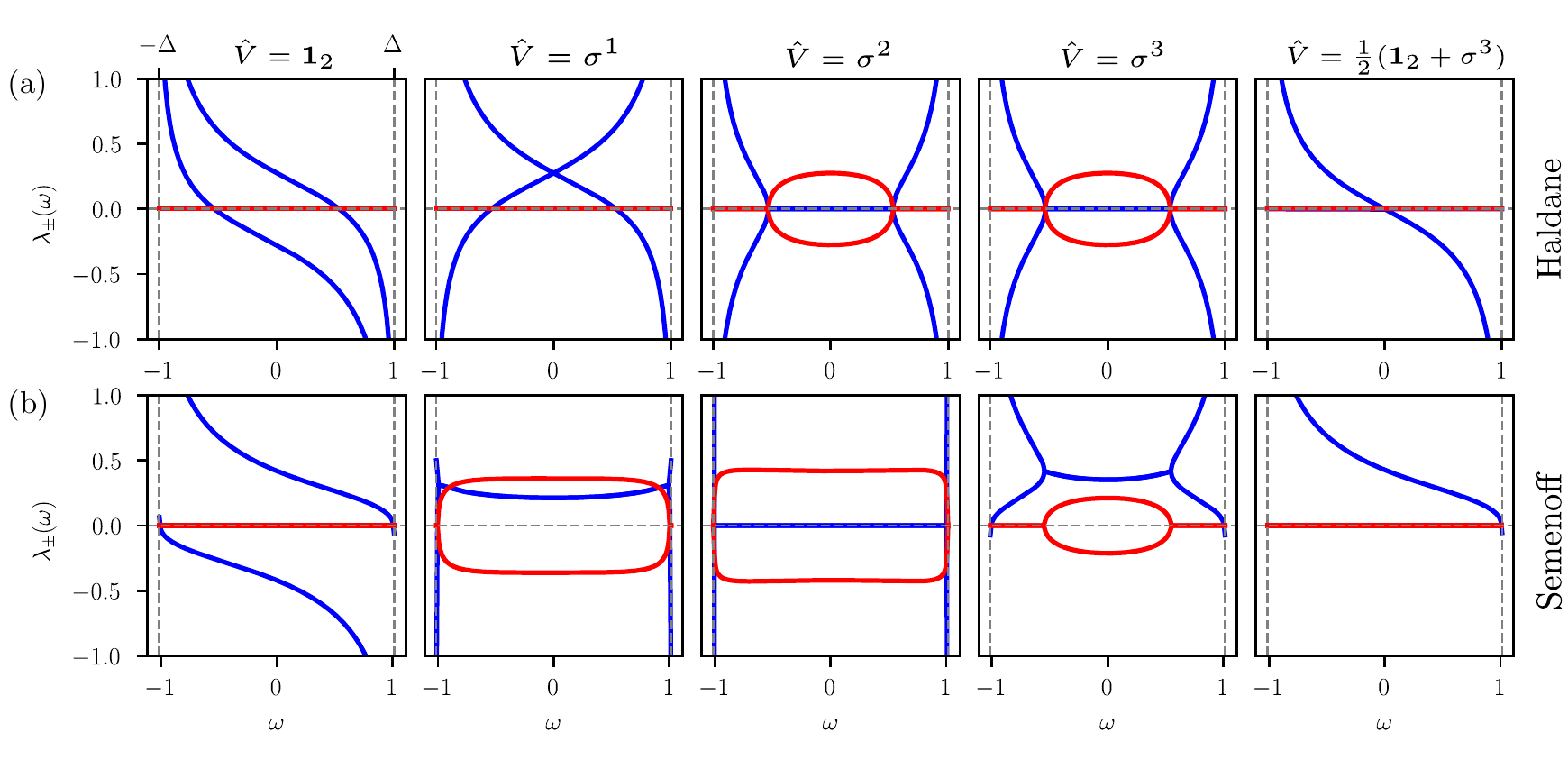}
\caption{
Eigenvalues $\lambda_{\pm}(\w)$ of $G(\omega)\hat{V}$ as a function of $\omega$ for the model \eqref{eq:haldane_hamiltonian}, shown for $\w$ in the gap region; the bulk gap is indicated by the vertical dashed lines. The columns correspond to impurities with different sublattice structure, i.e., different $\hat{V}$.
(a) Haldane insulator ($M=0, t'=0.2$).
(b) Semenoff insulator ($M=1, t'=0$).
Blue (red) lines represent the real (imaginary) part of $\lambda_{\pm}$.}
\label{fig:haldane_greens_function}
\end{center}
\end{figure*}
%%%%%%%%%%%%%%%%%%%%%%%%%%%%%%%%%%%%%%%%%%%%%%

The behavior in Fig.\,\ref{fig:haldane_impurity_spectra} is precisely in line with the prediction of Ref.\,\onlinecite{slager-15prb}. There it has been noted that the presence of in-gap bound states is connected to the behavior of the eigenvalues of the local Green's function in the absence of the impurity, $G(\w,\vec{r}=\vec{0})\equiv G(\omega)$, for energies $\w$ located in the gap. It has been shown that for a topologically trivial phase there {\it cannot} be a zero eigenvalue of $G(\w,\vec{r}=\vec{0})$ in the gap; moreover, in a topological phase there always {\it must} be at least one zero eigenvalue. This ensures the (non-)existence of bound states in the strong-impurity limit, $V_0\to\infty$. For arbitrary impurity strength there will be a bound state with energy $\omega$ if
\begin{equation} \label{eq:T_matrix_eq}
\det \left[\mathds{1}-V_0 \cdot G(\omega, \vec{r}=\vec{0}) \hat{V} \right]=0\ .
\end{equation}
By defining $\lambda_{\pm}(\omega)$ the two eigenvalues of $G(\omega)\hat{V}$, condition \eqref{eq:T_matrix_eq} is equivalent to solving the equation $\lambda_{\pm}(\omega) = 1/V_0$. In Fig.\,\ref{fig:haldane_greens_function} we show $\lambda_{\pm}(\omega)$ for the model \eqref{eq:haldane_hamiltonian} in both the topologically trivial and non-trivial phases and for different impurity types.
For both the site impurity and the real-valued bond impurity, the energies for which $\lambda_{\pm} = 0$ indeed correspond to the asymptotic ($V_0 \to\infty$) bound-state energies. If the eigenvalues $\lambda_{\pm}(\omega)$ have a finite imaginary part, a real-valued impurity cannot host a bound state at energy $\omega$. For example, in the case of the Semenoff insulator with a real bond impurity, the $\lambda_{\pm}$ are imaginary across the entire gap -- this agrees with the observed absence of bound states. We note that impurities of type $\mathbf{1}_{2}$ and $\sigma^{1}$ show identical behavior in the strong-$V_0$ limit w.r.t. the number and energies of the induced bound states; the same applies to $\sigma^{2}$ and $\sigma^{3}$ impurities.
Hence, the difference between the two phases is that the eigenvalues of $G(\omega)\hat{V}$ vanish in the gap for the topological phase -- leading to in-gap bound states -- but not for the trivial phase, as advocated and previously shown for the BHZ model \cite{slager-15prb}.

%%%%%%%%%%%%%%%%%%%%%%%%%%%%%%%%%%%%%%%%%%%%%%%%%%%%%%%%%%%%%%%%%%%%%%%%%%%

\section{Anisotropic hopping on the honeycomb lattice}
\label{sec:tatbtc-model}

\begin{figure}[!bt]
\begin{center}
\includegraphics[width=\columnwidth]{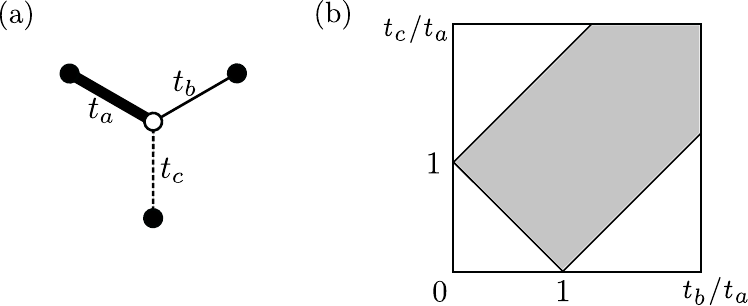}
\caption{
(a) Breaking of the discrete rotational symmetry of the honeycomb lattice due to anisotropic first-neighbor hoppings $t_a$, $t_b$ and $t_c$ as in model \eqref{eq:tatbtc}. (b) Corresponding phase diagram \cite{hasegawa-12prb}; the grey region is the semimetallic phase while white regions are insulating phases.
}
\label{fig:tatbtc}
\end{center}
\end{figure}

In this section we illustrate the impurity physics of the anisotropic tight-binding model on the honeycomb lattice governed by the Hamiltonian
\begin{equation}
\label{eq:tatbtc}
H_0 = - \sum_i \left( t_a c_i^\dag c_{i+\bs{\delta}_a}^\pd + t_b c_i^\dag c_{i+\bs{\delta}_b}^\pd  + t_c c_i^\dag c_{i+\bs{\delta}_c}^\pd  + {\rm H.c.}\,\right)\ .
\end{equation}
We restrict ourselves to first-neighbor hopping terms along the nearest-neighbor vectors $\bs{\delta}_j$, $j=a,b,c$, with distinct hopping amplitudes for each direction, $t_j$. The hopping anisotropy breaks the three-fold rotational symmetry $C_3$ of the honeycomb lattice, Fig.\,\ref{fig:tatbtc}\,(a), microscopically it may arise from applying uniaxial strain. In the low-energy spectrum, the anisotropy displaces the position of the Dirac cones in the Brillouin zone. If this displacement becomes sufficiently large, Dirac cones can pairwise merge and annihilate, \ie the bandstructure acquires an energy gap\,\cite{hasegawa-12prb}. The corresponding phase diagram is shown in Fig.\,\ref{fig:tatbtc}\,(b). In what follows, we concentrate on one of the gapped phases and choose $t_a>t_b>t_c$.

%%%%%%%%%%%%%%%%%%%%%%%%%%%%%%%%%%%%%%%%%%%%%%
%%%%%%%%%%%%%%%%%%%%%%%%%%%%%%%%%%%%%%%%%%%%%%

%%%%%%%%%%%%%%%%%%%%%%%%%%%%%%%%%%%%%%%%%%%%%%
\begin{figure*}[!bt]
\begin{center}
\includegraphics[width=0.92\textwidth]{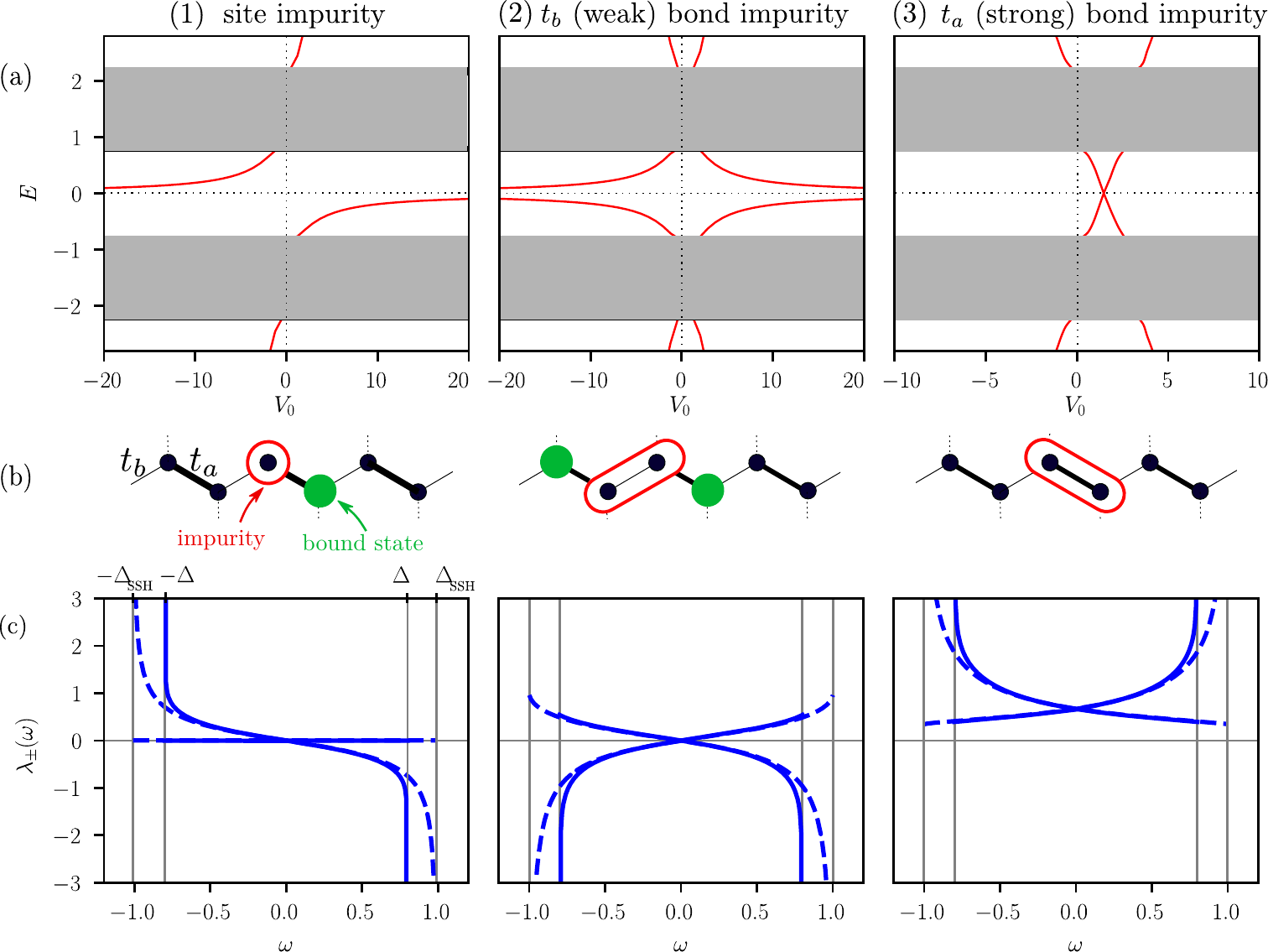}
\caption{
(a) Single-particle levels vs.\ impurity strength $V_0$ for the $t_a$--$t_b$--$t_c$ model \eqref{eq:tatbtc}, with $t_a=1.5$, $t_b=0.5$, $t_c=0.2$.
(b) Position of impurity and induced bound states, focussing on the chain formed by $t_a$ and $t_b$ hoppings.
(c) Real parts of the corresponding eigenvalues $\lambda_\pm(\omega)$ of $G(\omega)\hat{V}$ as a function of $\omega$ (solid); imaginary parts vanish everywhere inside the gap. The bulk gap $\Delta$ is indicated by the vertical lines.
Different columns show different impurity types, namely site impurity (left), bond impurity on $t_b$ bond (middle), and bond impurity on $t_a$ bond (right).
Panel (c) also shows the eigenvalues for the model with $t_c=0$, i.e., the SSH chain.
}
\label{fig:tatbtc_impurity_spectra}
\end{center}
\end{figure*}
%%%%%%%%%%%%%%%%%%%%%%%%%%%%%%%%%%%%%%%%%%%%%%

The energy spectrum of the anisotropic $t_a$--$t_b$--$t_c$ model \eqref{eq:tatbtc} in the presence of a site or bond impurity of strength $V_0$ is shown in Fig.\,\ref{fig:tatbtc_impurity_spectra}\,(a). Due to the broken $C_3$ rotation symmetry, the result for bond impurities depends on the orientation of the impurity bond (or, equivalently, the orientation of the defect unit cell). For a bond impurity on a weak $t_b$ bond, the energy of the bound states remains within the gap for strong $V_0$. A bond impurity on a $t_c$ bond is qualitatively the same as on a $t_b$ bond. On the contrary, if the impurity is on a strong $t_a$ bond, the in-gap states cross within the gap and then disappear into the bulk bands. A site impurity always gives rise to a bound state within the gap.

The above results can be rationalized in the limit $t_c\to0$, where the model reduces to dimerized chains, formed by the $t_a$ and $t_b$ bonds. The $t_a$--$t_b$ chain is equivalent to the Su--Schrieffer--Heeger (SSH) model, which is known to possess zero-energy edge states for an appropriate choice of boundaries\,\cite{heeger-88rmp781,su-79prl1698}. A site or bond impurity in the chain has codimension $1$ and thus behaves like a soft edge. By increasing the impurity strength we induce a new boundary to the system which may or may not bind an edge state, depending on the location of the impurity. This is exactly what we also see in the $t_a$--$t_b$--$t_c$ model, Figs.\,\ref{fig:tatbtc_impurity_spectra}\,(a,b).
We conclude that the impurity bound states of the $t_a$--$t_b$--$t_c$ model are \emph{inherited} from that of the SSH chain; this is supported by analyzing the eigenvalues $\lambda_{\pm}$ of $G\hat{V}$ for both the $t_a$--$t_b$--$t_c$ model and the SSH chain which are found to have the same structure, Fig.\,\ref{fig:tatbtc_impurity_spectra}\,(c).

%%%%%%%%%%%%%%%%%%%%%%%%%%%%%%%%%%%%%%%%%%%%%%
\begin{figure*}[tb!]
\begin{center}
\includegraphics[width=0.85\textwidth]{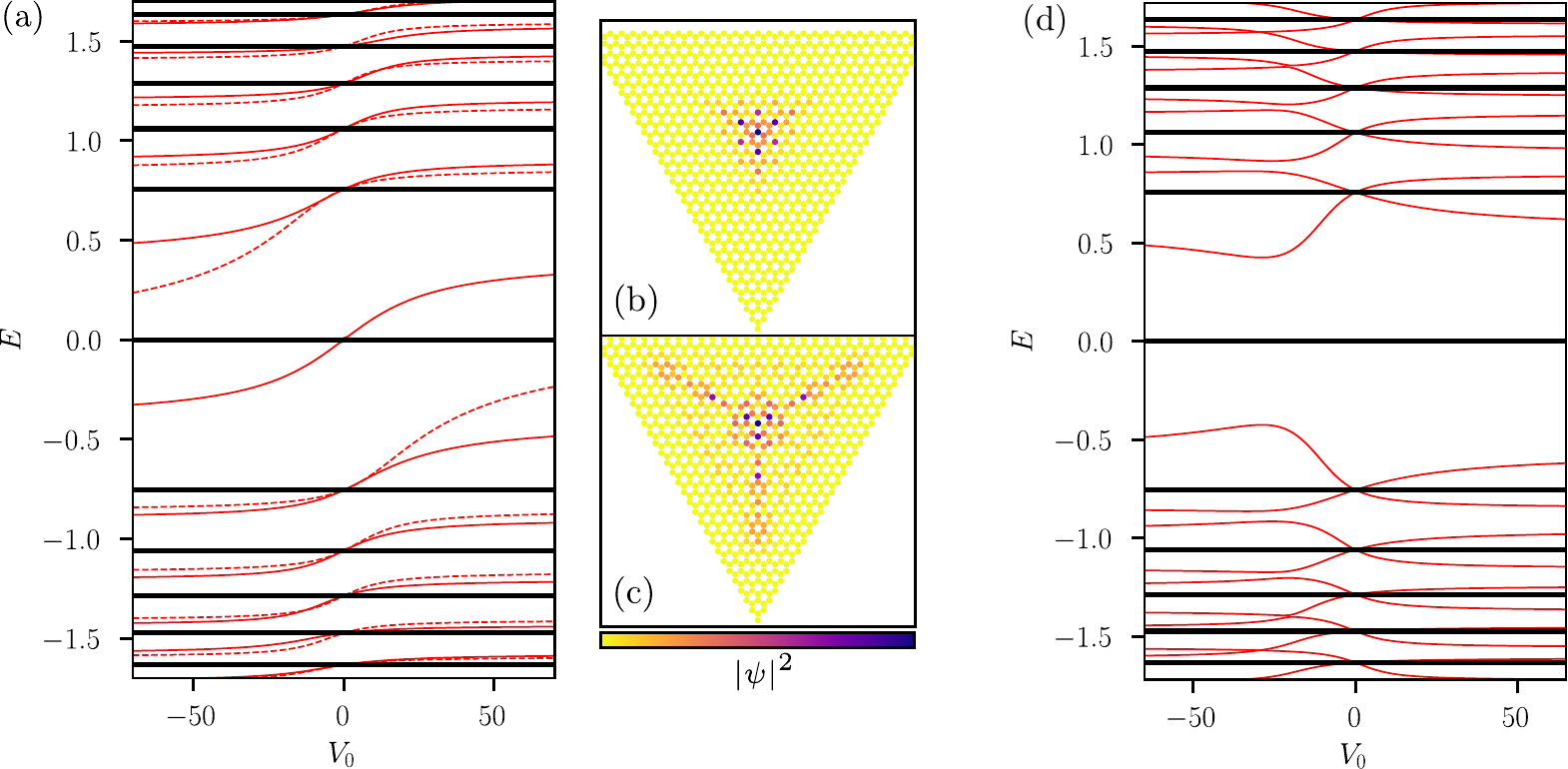}
\caption{
(a) Single-particle levels vs. impurity strength for the triaxially strained honeycomb lattice \eqref{hopping-strain}. Pseudo-Landau levels with energy $E_n^\pm =\pm (3/N)\sqrt{N^2 - n^2}$ are shown in black, here $N=31$. Red solid (dashed) lines show the energies of the site impurity when being located on sublattice A (sublattice B).
(b,c) Wavefunctions $|\psi|^2$ for in-gap bound states ($V_0=20$) closest to $E=0$, \ie the zeroth Landau level (b) and within the gap between third and fourth Landau level (c).
(d) Same as (a), but now for a bond impurity located in the center of the lattice.
}
\label{fig:exact_LLs_impurity}
\end{center}
\end{figure*}
%%%%%%%%%%%%%%%%%%%%%%%%%%%%%%%%%%%%%%%%%%%%%%

Let us consider the SSH chain in more detail, which is the paradigm not only of a one-dimensional symmetry protected topological phase\,\cite{footnote} but also of a model made out of alternating weak and strong bonds. For this model, the existence of edge states depends on the position of the boundary within the unit cell, and the situation with edge states is referred to as topological. However, in the bulk the topological and non-topological situations only differ by a shift of the unit cell by one site. The spectral response to a site impurity is insensitive to such a shift and hence does {\em not} detect whether or not the system displays edge states; the same applies to a bond impurity. Hence, the diagnostic of Ref.\,\onlinecite{slager-15prb} \emph{cannot} distinguish the situations with and without edge states; by continuity the same applies to the $t_a$--$t_b$--$t_c$ model.

Parenthetically, we recall the notion of a \emph{weak} topological phase, which refers to band topology inherited from a lower-dimensional system. Hence, the $t_a$--$t_b$--$t_c$ model is in a weak topological phase in the sense of the SSH chain. Moreover, it was recently shown that the gapped phases of the $t_a$--$t_b$--$t_c$ model can be interpreted as HOTI phases \cite{ezawa18prb045125}, displaying corner modes for appropriate system geometry. We will come back to the impurity response of HOTI models below in Sec.\,\ref{sec:HOTI}.

%%%%%%%%%%%%%%%%%%%%%%%%%%%%%%%%%%%%%%%%%%%%%%%%%%%%%%%%%%%%%%%%%%%%%%%%%%%

\section{Strain-induced pseudo-Landau levels}
\label{sec:PLL}

Inhomogeneous mechanical strain applied to a lattice system generically induces spatial variations in tight-binding hopping amplitudes. For graphene's honeycomb lattice, it has been shown that the effect of strain can be cast into a pseudo-vector potential appearing in the low-energy Dirac theory. In particular, a graphene flake subject to triaxial strain displays pseudo-Landau levels\,\cite{PhysRevLett.101.226804,guinea-10np30,levy-10s544,gomes-12n306,vozmediano-10pr109}.
In contrast to the case of Landau levels arising from a physical magnetic field, the strain-induced pseudo-magnetic field has opposite sign for the two valley momenta $K$ and $K'$, reflecting that strain preserves time-reversal symmetry. Time-reversal invariance also guarantees zero total Chern number, hence the resulting pseudo-Landau levels can be expected to be topologically trivial. \sr{Translational symmetry is, however, broken by the inhomogeneous strain field; a discrete three-fold rotational symmetry remains intact}.

A recent analysis in the context of a mechanically strained Kitaev spin liquid \cite{rachel-16prl-116} found that particular flux impurities placed in a triaxially strained honeycomb-lattice hopping model induce bound states inside the Landau-level gaps. According to the diagnostic of Ref.\,\onlinecite{slager-15prb}, this suggested that pseudo-Landau levels are to be classified as topological.

This motivates us to investigate the pseudo-Landau levels and their impurity response in more detail. Triaxial strain requires to work with open boundary conditions. In order to avoid complications arising from edge effects and imperfect pseudo-Landau levels away from zero energy, we focus on a particular limit of infinite electron-lattice coupling and maximum strain which has been shown to produce perfectly degenerate pseudo-Landau levels over the entire bandwidth for triangular-shaped systems\,\cite{rachel-16prl-117}. The corresponding tight-binding Hamiltonian is defined on a regular honeycomb lattice with inhomogeneous hopping,
\begin{equation}
H_0 = \sum_{i} \sum_{\alpha=1}^3 ( t_{i,\alpha}^N c_{i}^\dag c_{i+\bs{\delta}_\alpha}^\pd + {\rm H.c.} ) \,.
\end{equation}
% notation is now consistent with other models in this paper
Here the summation $i$ is over the sites of one sublattice (B), with the position $\bs{r}_i=0$ defining the center of the system. The hopping amplitudes are given by
\begin{equation}
\label{hopping-strain}
t_{i,\alpha}^N = \left(N-1-2 \bs{r}_i\cdot\bs{\delta}_\alpha\right)/N
\end{equation}
where $N\in\mathbb{N}$ specifies the linear system size, and the total number of sites is $N^2$. As shown in Ref.~\onlinecite{rachel-16prl-117}, the single-particle energies for this model can be obtained in closed form, with the result $E_n^\pm = \pm (3/N)\sqrt{N^2-n^2}$ with $n=0,1,\ldots, N$. The states at $E_n$ represent sharp pseudo-Landau levels; for $n\lesssim N$ they correspond to the low-energy pseudo-Landau levels obtained for weak strain in earlier work\,\cite{PhysRevLett.101.226804,guinea-10np30,vozmediano-10pr109}.

The single-particle spectrum in the presence of a single site impurity placed at the center of the lattice, $\bs{r}=0$, is shown in Fig.\,\ref{fig:exact_LLs_impurity}\,(a). An impurity on the B sublattice gives rise to an in-gap state in each of the gaps; the $V_0$ dependence reveals that each in-gap state derives from one of the pseudo-Landau levels. For an impurity on the A sublattice the same applies, with the exception of the $E=0$ Landau level which is sublattice-polarized and hence does not contribute a bound state here. 
Bound-state wavefunctions are illustrated in Fig.~\ref{fig:exact_LLs_impurity}\,(b,c); their localization length increases with decreasing gap size.
A bond impurity leads to two bound states in each gap, except for the lowest gap where the bound state deriving from the $E=0$ Landau level is missing as before, Fig.~\ref{fig:exact_LLs_impurity}(d).
In both Figs.~\ref{fig:exact_LLs_impurity}(a) and (d) the $V_0$ dependence of the in-gap states is very similar to that seen for the Haldane model, Fig.~\ref{fig:haldane_impurity_spectra}(a).
We note that deviating from the strong-strain limit of Ref.~\onlinecite{rachel-16prl-117} will introduce broadening of the finite-energy pseudo-Landau levels, such that (some of) the bound states may merge with the Landau levels, as reported in Ref.\,\onlinecite{rachel-16prl-116}.

%%%%%%%%%%%%%%%%%%%%%%%%%%%%%%%%%%%%%%%%%%%%%%
\begin{figure}[b!]
\begin{center}
\includegraphics[scale=1.0]{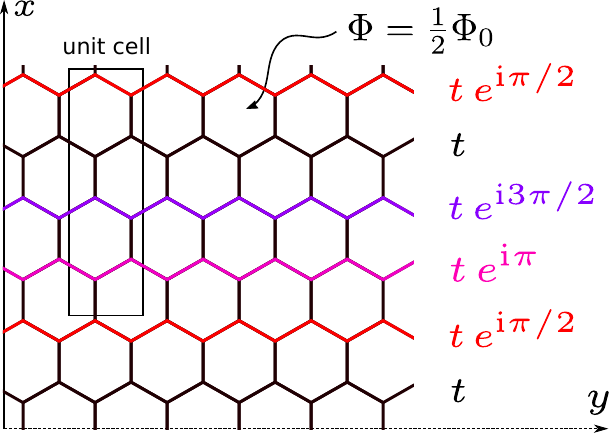}
\caption{
Pattern of hopping energies for the honeycomb-lattice Hofstadter model \eqref{eq:hofst} in Landau gauge with $\alpha = 1/2$. All vertical bonds have real hopping $t$.
}
\label{fig:hofstadter_model}
\end{center}
\end{figure}
%%%%%%%%%%%%%%%%%%%%%%%%%%%%%%%%%%%%%%%%%%%%%

Fig.\,\ref{fig:exact_LLs_impurity} show the presence of bound states in each gap for arbitrary impurity strength. As noted above, this suggests a state with non-trivial topology being realized, despite the Chern number being zero. We note that strain-induced modulations and corresponding pseudo-Landau levels were also studied in 1D and 3D lattices\,\cite{rachel-16prl-117}; they exhibit a similar response to an impurity (not shown here), and hence the same conclusion applies.

%%%%%%%%%%%%%%%%%%%%%%%%%%%%%%%%%%%%%%%%%%%%%%%%%%%%%%%%%%%%%%%%%%%%%%%%%%%

\section{Hofstadter bands}
\label{sec:hofstadter}

{In this section we study the case of a physical (orbital) magnetic field, applied perpendicular to the lattice, which leads to Landau levels and associated Hofstadter bands.
The square lattice version was first studied by Hofstadter\,\cite{hofstadter-76prb2239}; the honeycomb-lattice case has been discussed in Refs.\,\cite{bernevig-06ijmp,kohmoto-06prb235118,sato-08prb235322,agazzi-14jsp417,das-19arXiv1908.03483}.
The Hofstadter model is defined as a nearest-neighbor tight-binding model with additional Peierls phases,
\begin{equation}
\label{eq:hofst}
H_0 = -t \sum_{\langle i,j \rangle} c_{i}^{\dagger} \exp \left(i \frac{2\pi}{\Phi_0} \int_{j}^{i} \vec{A}\cdot d\vec{r} \right)  c_{j}^\pd + {\rm H.c.}
\end{equation}
with vector potential $\vec{A}$ and $\Phi_0 = h/e$ the Dirac flux quantum.
The integral is taken along a (linear) path from site $j$ to site $i$. In Landau gauge we have $\vec{A}(x,y) = \alpha \frac{\Phi_{0}}{\mathcal{A}} x \vec{e}_{y}$. The magnetic flux per honeycomb (of area $\mathcal{A}$) is then $\Phi = \alpha \Phi_0$. In Fig.\,\ref{fig:hofstadter_model} we show an example for the hopping amplitudes for $q\equiv1/\alpha=2$ in Landau gauge. We note that the doubling of the unit cell could be prevented by using the {\it optimal} gauge\,\cite{das-19arXiv1908.03483}; the results are of course independent of the gauge choice.

%%%%%%%%%%%%%%%%%%%%%%%%%%%%%%%%%%%%%%%%%
\begin{figure}[b!]
\begin{center}
\includegraphics[width=\columnwidth]{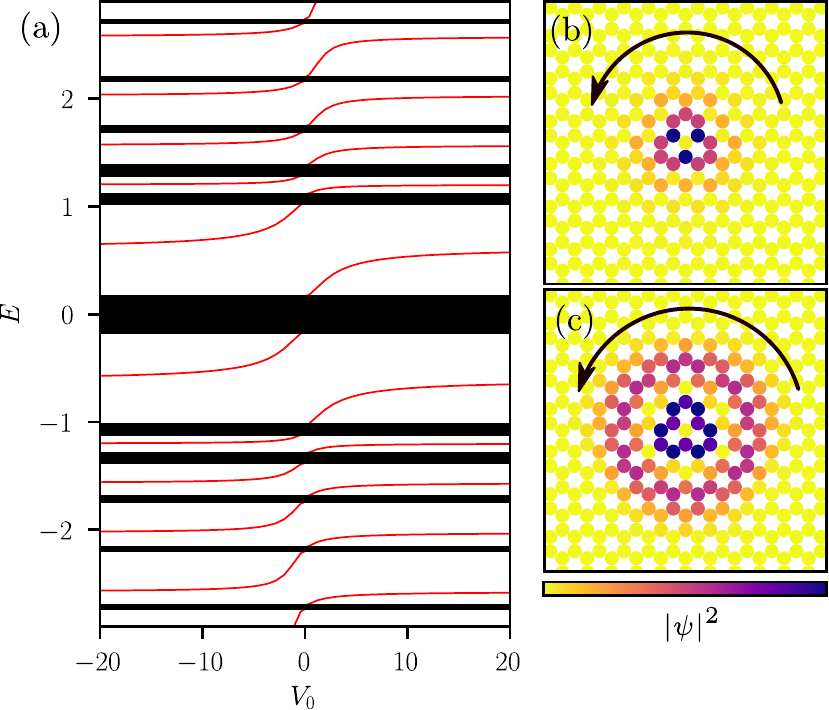}
\caption{
(a) Single-particle levels vs. impurity strength for the Hofstadter model \eqref{eq:hofst} with $\alpha = 1/6$. Red lines: in-gap states for a site impurity; black lines: Hofstadter bulk bands. (b, c) Wavefunction $|\psi|^2$ for the in-gap bound states ($V_0=20$) closest to $E=0$ (b) and within the gap with highest energy (c).
Black arrows show the direction of the current, \ie the chirality.
%(d) DOS and IPR  for an impurity strength $V_0=20$.  IPR is restricted to  nearest  and next-nearest neighbors of the site impurity. DOS and IPR are normalized by their maximum value. Each peak was broadened with a Lorentzian  of width $\gamma = 0.003$. Lattice with $1152$ sites and periodic boundary conditions imposed.
}
\label{fig:hofstadter_impurity_site}
\end{center}
\end{figure}
%%%%%%%%%%%%%%%%%%%%%%%%%%%%%%%%%%%%%%%%%

%%%%%%%%%%%%%%%%%%%%%%%%%%%%%%%%%%%%%%%%%%%%%
\begin{figure*}[t]
\begin{center}
\includegraphics{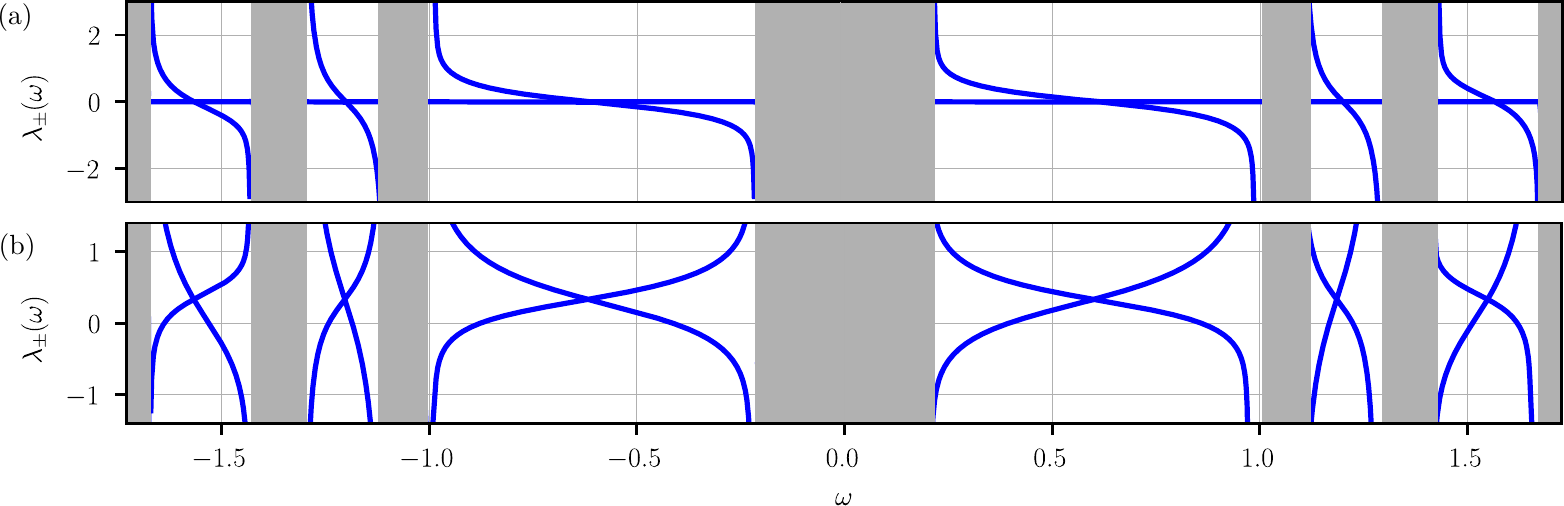}
\caption{
Real parts of the eigenvalues $\lambda_{\pm}(\omega)$ of $G(\omega)\hat{V}$ for the Hofstadter model \eqref{eq:hofst}.
(a) Site impurity $\hat{V} = \frac{1}{2}(1+\sigma^{3})$ and 
(b) bond impurity $\hat{V} = \sigma^{1}$.
}
\label{fig:hofstadter_gf}
\end{center}
\end{figure*}
%%%%%%%%%%%%%%%%%%%%%%%%%%%%%%%%%%%%%%%%%%%%%

In Fig.\,\ref{fig:hofstadter_impurity_site} we show the single-particle spectrum for $q=6$ in the presence of a site impurity. In general, the bulk band structure consists of $2q$ weakly dispersive bands (if $q\in\mathbb{N}$), resembling flat Landau levels near the band top and bottom and for large $q$.
Just like the previous results for the strained Landau levels, the spectra for a site impurity show a bound state in each gap. Moreover, the most localized in-gap state is the one whose energy lies in the largest energy gap (close to $|E|=0$), see Fig.\,\ref{fig:hofstadter_impurity_site}\,(b). For large $V_0$, all these states remain within the respective gap.
Again, the $V_0$ dependence of the bound-state energies in  Fig.\,\ref{fig:hofstadter_impurity_site}(a) is very similar to that of the Haldane model, Fig.~\ref{fig:haldane_impurity_spectra}(a). This also applies to a bond impurity (not shown).
The analysis is confirmed by the eigenvalues $\lambda_{\pm}(\omega)$ of $G(\omega)\hat{V}$, Fig.\,\ref{fig:hofstadter_gf}. Indeed, \emph{in each gap} the eigenvalues $\lambda_{\pm}$ resemble the eigenvalues obtained for the Haldane Hamiltonian.
Note that the impurity-bound states are chiral and have a current circulating around the impurity\,\cite{jha-17prb115434}, in accordance with the broken time-reversal symmetry. The chirality is indicated by the arrows in Fig.\,\ref{fig:hofstadter_impurity_site}\,(b,c) and the current can be explicitly calculated via $\bs{j} \sim \psi\vec\nabla\psi^\star - \psi^\star\vec\nabla\psi$.

As an aside, we note that for a real-valued bond impurity on a complex-valued bond of the Hofstadter model the impurity response is similar to that of an imaginary-valued bond impurity placed on a real-valued bond of the Haldane model. In particular, the behavior of in-gap bound states agrees in the limit of $V_0\to\infty$.

%%%%%%%%%%%%%%%%%%%%%%%%%%%%%%%%%%%%%%%%%%%%%%%%%%%%%%%%%%%%%%%%%%%%%%%%%%%

%%%%%%%%%%%%%%%%%%%%%%%%%%%%%%%%%%%%%%%%%%%%%
\begin{figure}[tb!]
\begin{center}
\includegraphics[width=\columnwidth]{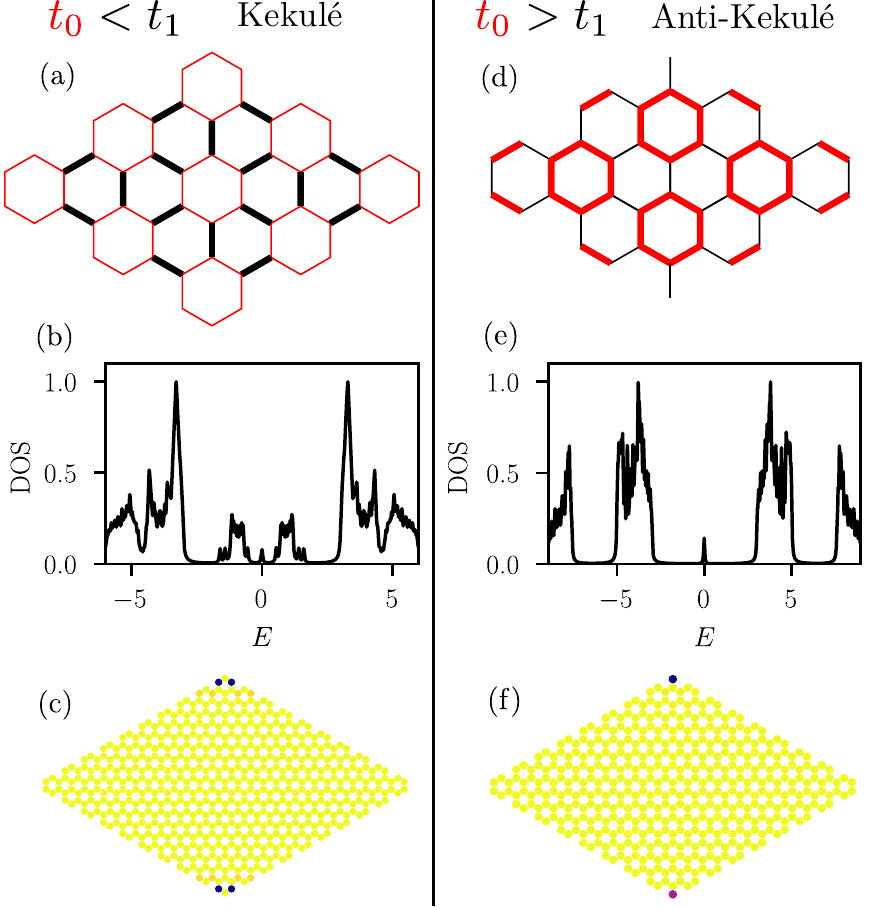}
\caption{(a)
Kekul\'e and (d) anti-Kekul\'e modulation of the honeycomb lattice. Red (black) bonds correspond to a hopping amplitude $t_0$ ($t_1$).
(b,e) Single-particle DOS of the Kekul\'e (anti-Kekul\'e) model computed on a lattice with 600 (434) sites with open boundaries. Corner states are visible at $E=0$. 
(c,f) Choices of the lattice geometry and its boundaries. Plots of $|\psi|^2$ of the zero-energy wavefunctions reveal themselves as corner modes; black (yellow) dots correspond to high (zero) intensity.}
\label{fig:HOTI_model}
\end{center}
\end{figure}
%%%%%%%%%%%%%%%%%%%%%%%%%%%%%%%%%%%%%%%%%%%%%

\section{Higher-order topological insulators on the honeycomb lattice}
\label{sec:HOTI}

Higher-order topological insulators have attracted much interest recently\,\cite{benalcazar-17s61,benalcazar-17prb,schindler-18saeaat0346,benalcazar-19prb,imhof-18np925}. These phases display boundary modes with co-dimension larger than 1 (\eg corner modes in 2D and hinge modes in 3D). Similar to the symmetry-protected quantized polarization of the SSH chain (leading to edge modes), HOTI models are characterized by a quantized higher multipole moment due to their crystalline symmetries \cite{benalcazar-17prb}.

%%%%%%%%%%%%%%%%%%%%%%%%%%%%%%%%%%%%%%%%%%%%%
\begin{figure*}[!tb]
\begin{center}
\includegraphics{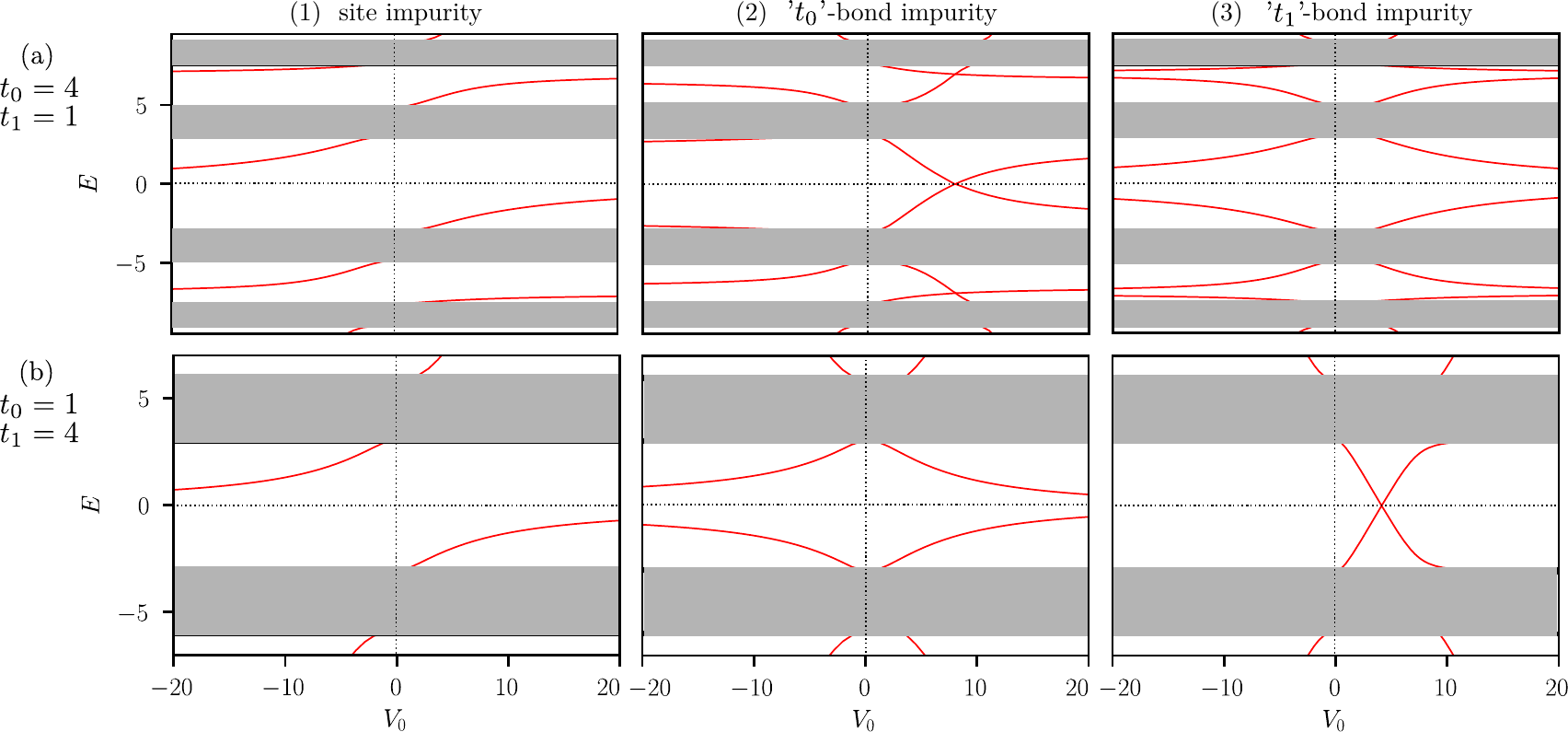}
\caption{
Single-particle levels vs.\ impurity strength for the HOTI model \eqref{eq:HOTI} (a) for anti-Kekul\'e modulation and (b) for Kekul\'e modulation.
Different columns show different impurity types, namely site impurity (left), bond impurity on $t_0$ bond (middle), and bond impurity on $t_1$ bond (right).
}
\label{fig:HOTI_impurity}
\end{center}
\end{figure*}
%%%%%%%%%%%%%%%%%%%%%%%%%%%%%%%%%%%%%%%%%%%%%

On the honeycomb lattice, HOTI phases have been discussed for Kekul\'e and anti-Kekul\'e  modulation patterns of the hopping amplitudes\,\cite{liu-19prl,mizoguchi-19arXiv1906.07928,wu-16sr,lee-19arXiv1903.02737,zangeneh-19arXiv1906.06605}. Here we follow Ref.\,\onlinecite{mizoguchi-19arXiv1906.07928} and consider the Hamiltonian
\begin{equation}
\label{eq:HOTI}
H_0 = -t_0 \sum_{\langle ij\rangle \,\in ~{\color{red}\varhexagon}} c_i^\dag c_j^\pd ~- t_1 \sum_{\langle ij \rangle \,\in \{ \shortmid\!, \medslash, \medbackslash\}} c_i^\dag c_j^\pd  ~+~ {\rm H.c.}\ .    % MnSymbol, marvosym, wasysym
\end{equation}
This pattern corresponds to hexagonal cells with intracell hopping $t_0$ and intercell hopping $t_1$. $t_0 < t_1$ is called Kekul\'e modulation and $t_0 > t_1$ \ anti-Kekul\'e modulation, see Fig.\,\ref{fig:HOTI_model}\,(a,d). For $t_0 = t_1$ the gapless semimetal is recovered, otherwise the system is insulating\,\cite{wu-12prb205102,mizoguchi-19arXiv1906.07928}, as can be seen from the single-particle density of states (DOS) plotted in Fig\,\ref{fig:HOTI_model}\,(b,e). It has been revealed that these two phases are both HOTIs with different topological properties, the Kekul\'e or dimer phase is characterised by a $\mathbb{Z}_2$ Berry phase and the anti-Kekul\'e or hexamer phase by a $\mathbb{Z}_6$ Berry phase\,\cite{mizoguchi-19arXiv1906.07928}. With open boundary conditions corner states appear for different choices of the boundary shape, Fig.\,\ref{fig:HOTI_model}\,(c,f).

%%%%%%%%%%%%%%%%%%%%%%%%%%%%%%%%%%%%%%%%%%%%%
\begin{figure}[!tb]
\begin{center}
\includegraphics{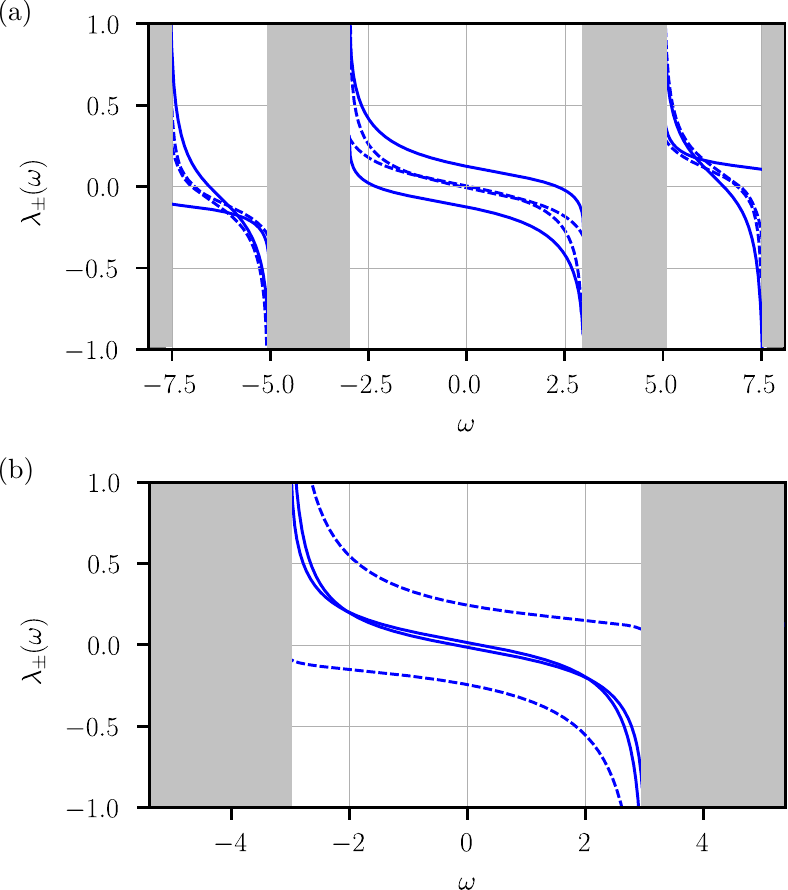}
\caption{
Real parts of the eigenvalues $\lambda_{\pm}(\omega)$ of $G(\omega)\hat{V}$ for the HOTI model \eqref{eq:HOTI}, here for $\hat{V}=\mathbf{1}_{2}$, i.e., two neighboring site impurities of equal strength. (a) Anti-Kekul\'e phase ($t_0 = 4, t_1 = 1$). (b) Kekul\'e phase ($t_0=1, t_1 = 4$). The solid (dashed) lines correspond to the case where a $t_0$ ($t_1$) bond is between the two site impurities.
}
\label{fig:HOTI_greens_function}
\end{center}
\end{figure}
%%%%%%%%%%%%%%%%%%%%%%%%%%%%%%%%%%%%%%%%%%%%%

%%%%%%%%%%%%%%%%%%%%%%%%%%%%%%%%%%%%%%%%%%%%%%%
\begin{figure}[!tb]
\centering
\includegraphics[scale=0.62]{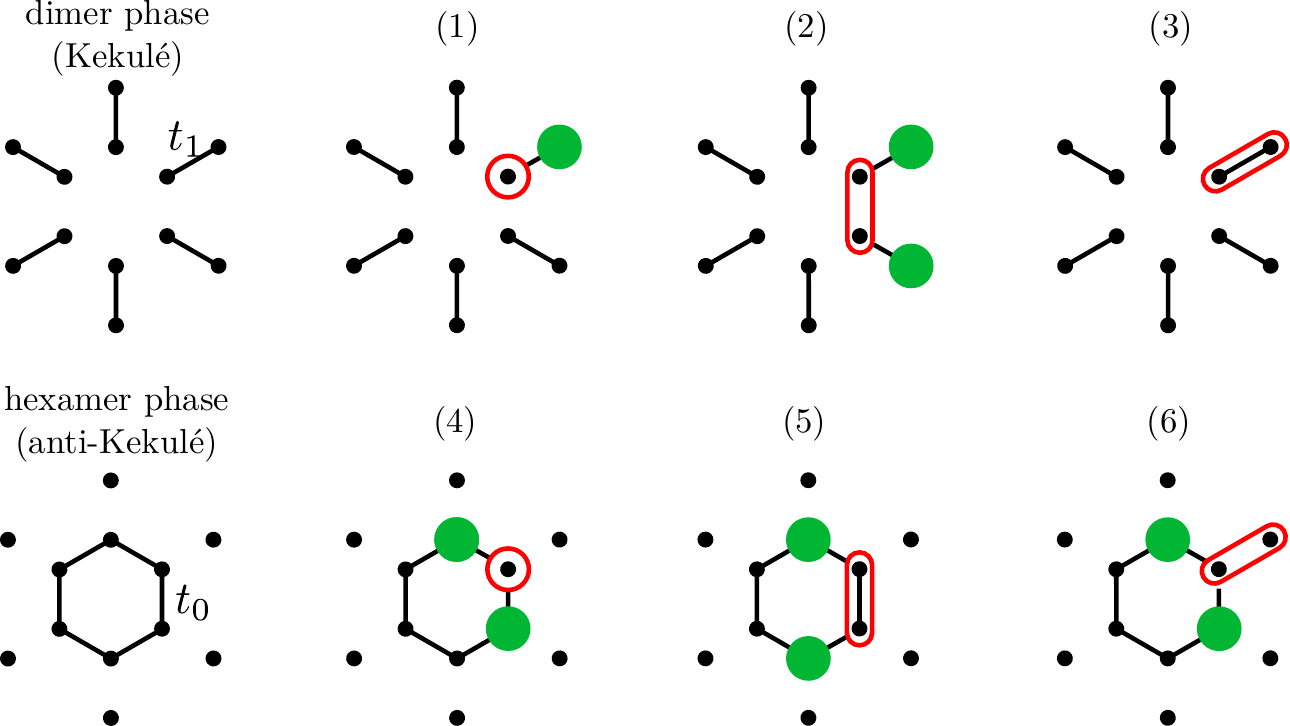}
\caption{
Sketches of decoupled \sr{oligomers}, \ie dimers and hexamers, obtained in the limits $t_0/t_1\to0$ or $t_0/t_1\to\infty$, respectively. Red circles indicate the position of a strong site or bond impurity, and green dots the position of the resulting bound states. Note that there will only be an in-gap bound state in the limit $V_0\to\infty$ if the resulting \sr{oligomer} contains an odd number of sites.
}
\label{fig:simplex}
\end{figure}
%%%%%%%%%%%%%%%%%%%%%%%%%%%%%%%%%%%%%%%%%%%%%%%

The single-particle energies of the HOTI model in the presence of an impurity (and with periodic boundary conditions) are shown in Fig.\,\ref{fig:HOTI_impurity}.
The spectra obtained for the Kekul\'e modulation, Fig.\,\ref{fig:HOTI_impurity}\,(b), are very similar to those obtained for the $t_a$--$t_b$--$t_c$ model, \cf Fig.\,\ref{fig:tatbtc_impurity_spectra}\,(a). They can be rationalized by considering the limit of decoupled dimers, $t_0/t_1\to 0$, Fig.\,\ref{fig:simplex}. A site impurity with large $V_0$ creates an isolated monomer, resulting in a low-energy impurity bound state. Similarly, a $t_0$-type bond impurity (between two dimers) creates two isolated monomers. In contrast, a $t_1$-type bond impurity influences an entire dimer, such that no low-energy state emerges for strong $V_0$. We note that low-energy bound states do emerge for $V_0 \approx t_1$, i.e., when the defect bond has small net hopping amplitude.
Similar to the case of the $t_a$--$t_b$--$t_c$ model, we see that the impurity response is not ``universal'', in the sense that both site and $t_0$ bond impurities always produce in-gap bound states, while this does not apply to $t_1$ bond impurities.

To rationalize the results for anti-Kekul\'e modulation, Fig.\,\ref{fig:HOTI_impurity}\,(a), we can proceed similarly. We adiabatically connect the hexamer phase to the opposite limit $t_1/t_0\to 0$ where the hexamers are fully decoupled, Fig.\,\ref{fig:simplex}. A site impurity creates a pentamer giving a zero-energy state (due to the odd number of remaining sites). A $t_1$-type bond impurity creates two pentamers. A $t_0$-type bond impurity creates a tetramer, which has two in-gap states (not at zero energy though). In contrast to the Kekul\'e phase, regardless of site or bond impurity and regardless of the type of bond we find in-gap bound states, also in the large-$V_0$ limit.

In summary, we can understand the impurity responses in each HOTI phase in the limit of decoupled \sr{oligomers}; the topology of the respective phase remains preserved upon taking this limit. When adding an impurity to either a dimer or a hexamer, we change the nature of this elementary \sr{oligomer} and its eigenenergies. This provides an intuitive explanation for the behavior of the energy response to an impurity, at least in the strong-$V_0$ limit. Using this approach, one can also engineer boundary shapes that give rise to corner modes in either of the phases, see Fig.\,\ref{fig:HOTI_model}\,(c) and (f).
Finally, we have calculated the eigenvalues $\lambda_\pm(\omega)$ of the local Green's function $G(\omega)\hat{V}$ for the anti-Kekul\'e and Kekul\'e phases. The $\omega$ dependence is essentially identical to the one of the $t_a$--$t_b$--$t_c$ model for site and bond impurity. In Fig.\,\ref{fig:HOTI_greens_function} we display results for two adjacent site impurities, \ie $\hat V =\mathds{1}_2$, indicating bound-state formation in all cases except for the Kekul\'e case with impurity on a $t_1$ bond, consistent with Fig.\,\ref{fig:simplex}.

%%%%%%%%%%%%%%%%%%%%%%%%%%%%%%%%%%%%%%%%%%%%%%%%%%%%%%%%%%%%%%%%%%%%%%%%%%%
%%%%%%%%%%%%%%%%%%%%%%%%%%%%%%%%%%%%%%%%%%%%%%%%%%%%%%%%%%%%%%%%%%%%%%%%%%%
%%%%%%%%%%%%%%%%%%%%%%%%%%%%%%%%%%%%%%%%%%%%%%%%%%%%%%%%%%%%%%%%%%%%%%%%%%%

\section{Discussion}
\label{sec:discussion}

In this section, we systematize the insights gained from the various models studied in Secs.~\ref{sec:haldane}-\ref{sec:HOTI}.

\subsection{Consistent diagnostics}

For the Haldane and Semenoff insulators as well as for the Hofstadter model, we have confirmed the diagnostic proposed in Ref.~\onlinecite{slager-15prb}, namely that the nature -- topological vs. non-topological -- of the phase can be read off from the spectral behavior of the in-gap bound states. Concretely, for a topological phase, bound states always exist in the gap for impurity strength $V_0\to\infty$ independent of the type of impurity, and these bound states remain trapped within the gap for (almost) all $V_0$. In contrast, for a non-topological phase, in-gap bound states only occur for a small interval of $V_0$ and not for $V_0\to\infty$.

For the Haldane and Semenoff insulators our results correspond to those for the BHZ model in Ref.~\onlinecite{slager-15prb}. Based on our analysis for the Hofstadter model, we conclude that the diagnostic yields a correct positive answer for systems with finite Chern number (including $Z_2$ topological insulators) and a correct negative answer for topologically trivial systems (with hoppings preserving the lattice symmetry, see below).

\subsection{Inconsistent diagnostics and false positives}

For the other models considered, we cannot confirm the diagnostic proposed in Ref.~\onlinecite{slager-15prb}. All of these models preserve time-reversal symmetry, implying a zero Chern number, and since we study spinless fermions any non-trivial topology would be expected to rely on crystalline symmetries.
The problems are twofold:

(i) The diagnostic delivers inconsistent results in the sense that, for some types (i.e. sublattice structures) of the impurity, bound states do indeed occur for (almost) all $V_0$, while for other types of impurities bound states only occur for a restricted range of $V_0$. This applies both to the anisotropic $t_a$--$t_b$--$t_c$ model and to the Kekul\'e-modulated HOTI phase.

(ii) The diagnostic delivers false positive results, i.e., impurity bound states occur for (almost) all $V_0$ in a phase which is non-topological or for reasons which are unrelated to topology. The simplest example is the SSH model where the impurity response is {\it identical} in the trivial and in the topological phase. Here (but also for the HOTI models) the existence of bound states can be traced to the behavior of isolated \sr{oligomers} and is therefore unrelated to (higher-order) topology. A false positive also occurs for the triaxially strained honeycomb lattice: Its impurity response is essentially identical to the one of the Hofstadter model, but while the latter has a finite Chern number, the former does not.

A common aspect of these models is that the hopping pattern breaks the lattice symmetry, i.e., the model consists of strong and weak bonds. We are forced to conclude that the diagnostic of Ref.~\onlinecite{slager-15prb} fails for such models. 
%Consistent with this, we have seen that the spectral response to impurities placed into an SSH chain -- another model with strong and weak bonds -- is unable to detect whether or not the system displays boundary states.

For completeness, we mention that we have also looked into other HOTI models: On the square lattice, a model consisting of strong and weak bonds \cite{benalcazar-17s61,benalcazar-17prb} was shown to realize a topologically trivial state in the absence of applied magnetic flux, but realizes a HOTI phase when subject to $\pi$ flux. For both cases we have determined the spectral response to impurities, and we find results (not shown) qualitatively similar to that for the honeycomb lattice HOTI models, regardless of whether or not a HOTI phase is realized (\ie whether zero flux or $\pi$ flux is applied).

\subsection{Experimental verification}

There are several avenues to realize local impurities in two-dimensional systems and to verify our predictions.
In artificially engineered systems such as semiconductor heterostructures, the role of the impurity can be taken by a local gate, implemented next to or below an effective lattice site, with the gate voltage taking the role of $V_0$. Local spectroscopic measurements are conveniently performed by means of scanning tunneling spectroscopy. Alternatively, an electrostatic potential applied by an STM tip can also introduce an impurity in a surface system. Hence, we believe that measuring bound-state energies as function of $V_0$ is indeed possible, with large-gap honeycomb topological insulators such as bismuthene\,\cite{reis-17s287} providing a perfect testing ground.

In addition, we note that there have been several breakthroughs in realizing topological states of matter such as 1D and 2D versions of the Su--Schrieffer--Heeger model\,\cite{heeger-88rmp781,su-79prl1698,drost-17np668} and of higher-order topological insulators\,\cite{benalcazar-17prb,benalcazar-17s61}, \eg on the kagome lattice\,\cite{kempkes-19arXiv1905.06053}. Also in topolectrical circuits\,\cite{imhof-18np925,zangeneh-19arXiv1906.06605} it is straightforward to integrate tunable impurities.

An immediate question is to which extent the effect of single intentional impurities can be distinguished from that of generic disorder present in condensed matter systems. To this end we have checked the robustness of the results against disorder in the cases of the Haldane and Semenoff models. We find that the dominant effect of disorder that is weak compared to the gap size is to smear out the edges of the gap. If the bound-state energy is very close to the band edge this might complicate the detection of the impurity-induced bound state. In the models at hand, this is the case for weak impurity potential, $|V_0|<1$, see Fig.\,\ref{fig:haldane_impurity_spectra}. We conclude that the results of the paper still hold provided that the disorder strength is small compared to the size of the bulk gap.

\subsection{Outlook}

Future work can extend our analysis in various directions. Clearly, more studies of HOTI models are in order, given that the work of Ref.\,\onlinecite{slager-15prb} was geared at first-order topological phases.
Consequently, it might be interesting to test whether statements about codimension-$1$ and codimension-$2$ impurities would be replaced by codimension $n$ and codimension $n+1$ impurities for an $n$th order topological insulators. In fact, at least the results for the Kekul\'e phase, Fig.\,\ref{fig:HOTI_greens_function}\,(b), seem to suggest precisely that. In this figure, we had chosen an impurity potential $\hat V \sim \mathds{1}_2$ such that we can quantitatively compare with Ref.\,\onlinecite{slager-15prb}; and indeed the codimension-$2$ impurity for the second-order tooplogical phase behaves just like the codimension-$1$ impurity case shown in Ref.\,\onlinecite{slager-15prb}.
For HOTI phases it will also be interesting to study the influence of the impurity shape in order to test sensitivity w.r.t. the quadrupole moment.

An entirely different, but experimentally highly relevant issue is to extend this line of research to interacting versions of topological insulators\,\cite{rachel18rpp116501,hohenadler-13jpcm143201,rachel-12prb075106}.

%%%%%%%%%%%%%%%%%%%%%%%%%%%%%%%%%%%%%%%%%%%%%%%%%%%%%%%%%%%%%%%%%%%%%%%%%%%
%%%%%%%%%%%%%%%%%%%%%%%%%%%%%%%%%%%%%%%%%%%%%%%%%%%%%%%%%%%%%%%%%%%%%%%%%%%
%%%%%%%%%%%%%%%%%%%%%%%%%%%%%%%%%%%%%%%%%%%%%%%%%%%%%%%%%%%%%%%%%%%%%%%%%%%

\section{Summary}
\label{sec:summ}

In summary, we have investigated several honeycomb lattice models with different topologies and different symmetries which feature gaps in their energy spectrum. Motivated by the proposal \cite{slager-15prb} to diagnose non-trivial topology via point-like impurities, we have studied the spectral response with respect to an impurity potential. As advocated in Ref.\,\onlinecite{slager-15prb}, topologically non-trivial phases feature in-gap bound states for arbitrary impurity strength, while trivial phases do not -- at best, they display in-gap bound states for fine-tuned values of the impurity strength.

We have found the diagnostic to work for the Haldane model, a Chern insulator with broken time-reversal symmetry, and the Semenoff model, a trivial insulator with broken inversion symmetry, as well as for the Hofstadter model, a lattice version of the quantum Hall effect.
In contrast, for several models whose electronic properties are determined by a pattern of weak and strong bonds, the impurity response cannot distinguish between topological trivial and non-trivial phases: For some or all type of impurities, the diagnostic incorrectly suggests topologically non-trivial behavior. Prominent examples are the anisotropic $t_a$--$t_b$--$t_c$ model and the triaxially strained hopping model of the honeycomb lattice.
We conclude that the diagnostic in terms of the bound-state response to a local impurity works only for models preserving the lattice symmetries, but not if rotation or translation symmetry are broken as the essential ingredient for the electronic properties of the model.

%%%%%%%%%%%%%%%%%%%%%%%%%%%%%%%%%%%%%%%%%%%%%%%%%%%%%%%%%%%%%%%%%%%%%%%%%%%

\acknowledgments

We acknowledge discussions with D. P. Arovas, R.-J. Slager, and M. Maksymenko and earlier collaborations with D. P. Arovas and I. Goethel.
This research was supported in part by the National Science Foundation under Grant No.\ NSF PHY-1748958.
SR acknowledges hospitality from KITP St.\ Barbara and support from the Australian Research Council through Grants No.\ FT180100211 and DP200101118.
MV acknowledges support from the DFG through SFB 1143 (project-id 247310070) and the W\"urzburg-Dresden Cluster of Excellence on Complexity and Topology in Quantum Matter -- \textit{ct.qmat} (EXC 2147, project-id 39085490).

%%%%%%%%%%%%%%%%%%%%%%%%%%%%%%%%%%%%%%%%%%%%%%%%%%%%%%%%%%%%%%%%%%%%%%%%%%%

\bibliographystyle{prsty} %abbrvnat.bst} %plainnat.bst} %unsrt}
\bibliography{impurity_TI}

\begin{thebibliography}{10}
\bibitem{footnote}
Note that in 1D only atomic and obstructed atomic insulators
  exist\,\cite{bradlyn-17n298,marzari-12rmp1419,alexandradinata-14prb155114,hwang-19arXiv},
  thus the term ``topological phase'' in 1D refers to the latter.
\end{thebibliography}

\end{document}